\documentclass[prc,superscriptaddress,unsortedaddress,twocolumn,preprintnumbers,amsmath,amssymb,floatfix,dvipdfmx]{revtex4}
\usepackage{amsmath}
\usepackage{amssymb}
\usepackage{mathrsfs}
\usepackage{bm}
\usepackage{color}
\usepackage{graphicx}
\usepackage{braket}
\usepackage{mathrsfs}
\usepackage{ulem}
\usepackage{here}

\makeatletter
\let\MYcaption\@makecaption
\makeatother
\usepackage{subcaption}
\makeatletter
\let\@makecaption\MYcaption
\makeatother

\makeatletter
\newcommand{\figcaption}[1]{\def\@captype{figure}\caption{#1}}
\newcommand{\tblcaption}[1]{\def\@captype{table}\caption{#1}}
\makeatother

\begin{document}

\title{
Systematic analysis of $t$ and $^3$He breakup reactions
}

\author{Shoya Ogawa}
\email[]{s-ogawa@phys.kyushu-u.ac.jp}
\affiliation{Department of Physics, Kyushu University, Fukuoka 819-0395, Japan}

\author{Shin Watanabe}
\affiliation{National Institute of Technology, Gifu College, Gifu 501-0495, Japan}
\affiliation{RIKEN, Nishina Center, Wako, Saitama 351-0198, Japan}

\author{Takuma Matsumoto}
\affiliation{Department of Physics, Kyushu University, Fukuoka 819-0395, Japan}
\affiliation{Research Center for Nuclear Physics (RCNP), Osaka University, Ibaraki 567-0047, Japan}

\author{Kazuyuki Ogata}
\affiliation{Department of Physics, Kyushu University, Fukuoka 819-0395, Japan}
\affiliation{Research Center for Nuclear Physics (RCNP), Osaka University, Ibaraki 567-0047, Japan}

\date{\today}

\begin{abstract}
 \noindent
 {\bf Background}: 
 Systematic measurement of $t$ and $^3$He knockout processes is planned. 
 The weakly-bound nature of these nuclei may affect the interpretation of 
 forthcoming knockout reaction data.
 \\
 {\bf Purpose}: 
 We aim at clarifying breakup properties of $t$ and $^3$He by investigating 
 their elastic and breakup cross sections.
 \\
 {\bf Methods}: 
 We employ the four-body continuum-discretized coupled-channels method with 
 the eikonal approximation to describe the $t$ and $^3$He 
 reactions.
 \\
 {\bf Results}: 
 The breakup cross section of $t$ is found to be almost the same as that 
 of $^3$He and is about one-third of that of $d$. 
 Coulomb breakup plays negligible role in the breakup of $t$ and $^3$He,
 in contrast to in the deuteron breakup reaction.
 It is found that $t$ and $^3$He tend to breakup into three nucleons 
 rather than $d$ and a nucleon.
 \\
 {\bf Conclusions}:
 It is shown that the breakup cross sections of $t$ and $^3$He are not 
 as large as those of $d$ but non-negligible. Because about 80\% of them 
 corresponds to the three-nucleon breakup process, 
 a four-body breakup reaction model is necessary to quantitatively 
 describe the breakup of $t$ and $^3$He.
\end{abstract}

\maketitle

\section{Introduction}

It is quite well known that $\alpha$ cluster states appear in light-mass nuclei. 
Recently, motivated by the theoretical prediction by Typel~\cite{Typel10} 
and its experimental confirmation~\cite{Tanaka21},
existence of $\alpha$ in medium-heavy nuclei has become a hot subject 
in nuclear physics~\cite{Yoshida16,Yoshida22}. 
Furthermore, the existence of $d$, $t$, and $^3$He are going to be studied 
by cluster knockout reactions.
Investigation of the $t$ and $^3$He cluster states of nuclei,
their neutron and proton number dependence in particular, is considered to be crucial for
determining  the symmetry energy term in the equation of state~\cite{Gaidarov21,Chen04,Ono14}.
However, there are very few theoretical studies of $t$ and $^3$He clusters.
In addition, 
despite the binding energies of $t$ and $^3$He are only about 2 MeV/nucleon,
their breakup effects on reaction observables have not been clarified well.
Under the circumstance that $t$ and $^3$He knockout reactions are going to be systematically measured,
it will be important clarify the breakup property of $t$ and $^3$He.

In Ref.~\cite{Iseri86}, the $^3$He breakup reaction was investigated,
in which $^3$He was treated as a $d+p$ two-body system. 
Because $d$ is fragile, however,
it is desirable to describe $^3$He as a $p+p+n$ three-body system;
including a target nucleus T, the reaction system consists of four particles.
The four-body continuum-discretized coupled-channels method 
(four-body CDCC)~\cite{Matsumoto04,Matsumoto06} is one of the best models for this purpose.

In this study, we investigate the four-body breakup reaction of $t$ and $^3$He
by using four-body CDCC to clarify the breakup properties of these nuclei
and understand their breakup mechanism due to the nuclear and Coulomb interactions.
Because coupled-channel calculations with the Coulomb breakup require high numerical costs
in general,
we use eikonal CDCC (E-CDCC)~\cite{Ogata03,Ogata09,Ogata10,Fukui12},
in which the coupled-channel calculations are performed with the eikonal approximation.
Using E-CDCC,
we can take into account the Coulomb breakup precisely with low computational cost.
We examine the description of the $^3$He breakup reaction with a $d+p$ two-body model, 
i.e., three-body CDCC.

This paper is organized as follows. In Sec. II, we describe the
theoretical framework. In Sec. III, we present and discuss the numerical results.
Finally, in Sec. IV, we give a summary of this study.

\section{Formalism}
\subsection{Eikonal CDCC}

In the four-body reaction system,
the Schr\"odinger equation is written as
\begin{eqnarray}
 \label{eq:4body-eq}
 \left[
  K_{R}
  + \sum_{i \in t{\rm ~or~ ^{3}He}} U_i
  + U_{\rm C} + h - E
 \right]
 \Psi(\bm{\xi},\bm{R}) = 0,
\end{eqnarray}
where $\bm{R}$ represents the coordinate between the target T and the center-of-mass (c.m.)
of the projectile.
The operator $K_{R}$ is the kinetic energy associated with $\bm{R}$,
$h$ is the internal Hamiltonian of the projectile,
and $\bm{\xi}$ is the intrinsic coordinate.
The optical potential between T and each nucleon in the projectile is denoted by $U_i$.
The Coulomb potential between a proton and T is represented by $U_{\rm C}$;
we investigate the effect of Coulomb breakup of $t$ and $^3$He in this study. 
In E-CDCC,
the scattering wave function is represented as
\begin{eqnarray}
 \label{eq:wf-ECDCC}
 \Psi(\bm{\xi},\bm{R})
  &=&
  \sum_{nIm} \psi_{nIm}(b,Z) 
  \Phi_{nIm}(\bm{\xi})  e^{iK_{n}Z} 
  \nonumber \\
 &&~~~~~~~~~~~~~~~~
  \times
  e^{i(m_{0}-m)\varphi_{R}}
  \phi^{\rm C}_{n}(R) ,
\end{eqnarray}
where $b$ is the impact parameter.
The position of the $Z$-axis and the azimuthal angle of $\bm{R}$ are denoted by
$Z$ and $\varphi_{R}$, respectively.
$\Phi_{nIm}$ is the $n$th discretized state of the projectile 
with the total spin $I$ and its projection on the $z$-axis $m$,
and $m_0$ is the $z$ component of the total spin of the ground state.
We denote $\gamma=\{n,I,m\}$ in this manuscript.
The wavenumber $K_n$ is written as
\begin{eqnarray}
 K_{n} = \frac{\sqrt{2\mu(E-\varepsilon_n)}}{\hbar},
\end{eqnarray}
where $\varepsilon_n$ is the eigen energy of $\Phi_{\gamma}$
and $\mu$ is the reduced mass between the projectile and T.
$\phi^{\rm C}_{n}$ in Eq.~\eqref{eq:wf-ECDCC} is 
the incident-wave part of the Coulomb wave function given by
\begin{eqnarray}
 \phi^{\rm C}_{n}(R) 
  =
  e^{i\eta_{n} \ln{[K_{n}R - K_{n}Z]}} 
\end{eqnarray}
with
\begin{eqnarray}
 \eta_{n} = \frac{Z_{\rm P}Z_{\rm T}e^2}{\hbar K_{n}} .
\end{eqnarray}
Here, $Z_{\rm P}$ and $Z_{\rm T}$ are the atomic numbers of 
the projectile and T, respectively.
Inserting Eq.~\eqref{eq:wf-ECDCC} into Eq.~\eqref{eq:4body-eq},
the following equation for $\psi_{\gamma}$ is obtained:
\begin{eqnarray}
 \label{eq:ECDCC-eq}
  &&\frac{\partial}{\partial Z} \psi_{\gamma}(b,Z)
  =
  \nonumber \\
 &&~~~~~~~~
  \frac{1}{i\hbar v_{n}}
  \sum_{\gamma'}
  \mathcal{F}_{\gamma\gamma'}(\bm{R}) \psi_{\gamma'}(b,z)
  e^{i(m-m')\varphi_{R}}
  \mathcal{R}_{\gamma\gamma'}(b,z) 
  \nonumber \\
\end{eqnarray}
with
\begin{eqnarray}
 \mathcal{R}_{\gamma\gamma'}(b,z)
  =
  \frac{(K_{n'}R - K_{n'}z)^{i\eta_{n'}}}{(K_{n}R - K_{n}z)^{i\eta_{n}}}
  e^{i(K_{n'} - K_{n})z} 
\end{eqnarray}
and
\begin{eqnarray}
 \mathcal{F}_{\gamma\gamma'}(\bm{R})
  =
  \braket{\Phi_{\gamma} | \textstyle\sum_{i \in {\rm P}}U_{i} | \Phi_{\gamma'}}_{\bm{\xi}} .
\end{eqnarray}
The subscript $\bm{\xi}$ of $\braket{\cdots}$ means the integral variable.

\subsection{Gaussian expansion method}

\begin{figure}[tbp]
 \centering
 \includegraphics[scale=0.25]{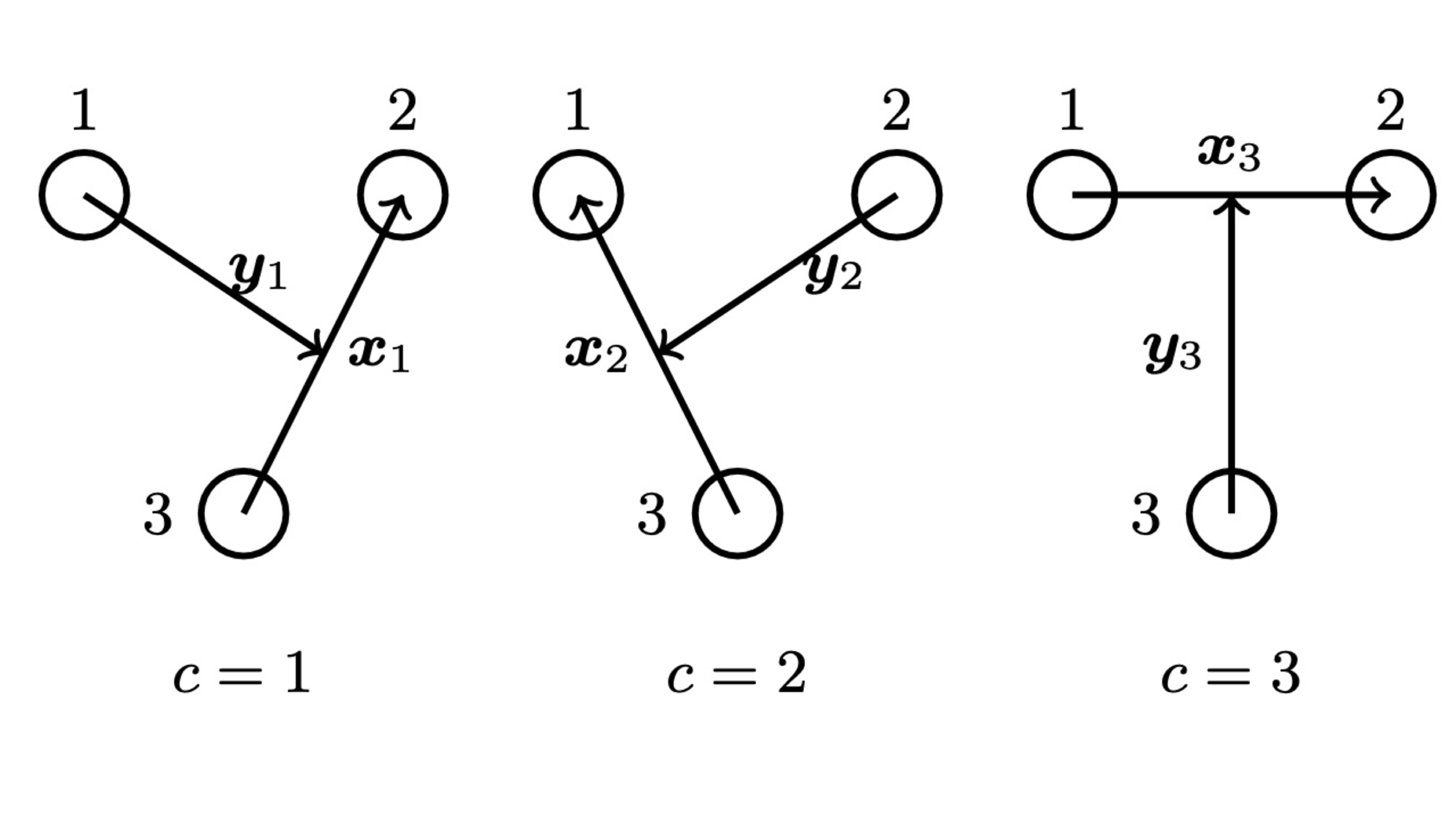}
 \caption{
 The Jacobi coordinate for the three-body system.
 Particles 1, 2, and 3 correspond to $n$, $n$, and $p$ 
 ($p$, $p$, and $n$) for $t$ ($^3$He), respectively.
 }
 \label{fig:jacobi}
\end{figure}

We apply the Gaussian expansion method (GEM)~\cite{Hiyama03} to obtain the ground 
and the dicretized-continuum states of $t$ and $^3$He.
In GEM,
a wave function of the three-body system is expanded with Gaussian basis 
on the Jacobi coordinate as shown in Fig.~\ref{fig:jacobi},
and the basis are described as
\begin{eqnarray}
 \phi_{i\lambda}(\bm{x}_c) 
  &=&
  x^{\lambda}_{c} e^{-(x/x_{i})^2} Y_{\lambda}(\Omega_{x_c}), 
  \\
 \varphi_{j\ell}(\bm{y}_c) 
  &=&
  y^{\ell}_{c} e^{-(y/y_{j})^2} Y_{\ell}(\Omega_{y_c}) 
\end{eqnarray}
with
\begin{eqnarray}
 x_{i} &=& (x_{\rm max}/x_{0})^{(i-1)/i_{\rm max}} ,
  \\
 y_{i} &=& (y_{\rm max}/y_{0})^{(j-1)/j_{\rm max}} .
\end{eqnarray}
Using the basis, 
we diagonalize the following Hamiltonian:
\begin{eqnarray}
 \label{eq:hamiltonian_t}
 h &=& K_{x} + K_{y} + V_{pn} + V_{pn} + V_{nn} 
\end{eqnarray}
for $t$, and 
\begin{eqnarray}
 \label{eq:hamiltonian_3He}
 h &=& K_{x} + K_{y} + V_{pn} + V_{pn} + V_{pp} + V_{\rm C} 
\end{eqnarray}
for $^3$He.
Here, $K_x$ ($K_y$) means the kinetic energy operator associated with $\bm{x}$ ($\bm{y}$).
The interactions for the $p$-$p$, $n$-$n$, and $p$-$n$ systems are represented as
$V_{pp}$, $V_{nn}$, and $V_{pn}$, respectively.
In Eq.~\eqref{eq:hamiltonian_3He}, 
$V_{\rm C}$ is the Coulomb interaction between the two protons.
%

\section{Results and Discussion}
\subsection{Three-body model for $t$ and $^3$He}

First, we obtain the ground-state wave functions of $t$ and $^3$He by using GEM.
In this study, we adopt the nucleon-nucleon Minnesota interaction~\cite{Thompson77}.
We neglect the spin of each nucleon for simplicity.
Thus, we use the ($S,T$) = ($0,1$) component of the Minnesota interaction for $V_{pp}$ 
and $V_{nn}$, where $S$ ($T$) is the total spin (isospin) of the two nucleons,
whereas we use the ($S,T$) = ($1,0$) component for $V_{pn}$.
To reproduce the binding energies of $t$ and $^3$He,
a phenomenological three-body interaction
\begin{eqnarray}
 V_{3b}(x,y) = V_{3} e^{-\nu(x^2+y^2)}
\end{eqnarray}
is added to $h$ of $t$ and $^3$He.
In the present analysis, $V_3=9.7$~MeV and  $\nu=0.1$~$\text{fm}^{-2}$.
The parameter sets of the Gaussian basis are common in both the $t$ and $^3$He calculations, 
and summarized in TABLE~\ref{tab:parameters-GEM}.
The spin-parity $I^{\pi}$ for the ground states is $0^+$ because
we neglect the spin of each nucleon in $t$ and $^3$He.
The results of the ground-state energies and root-mean-square radii
are shown in TABLE~\ref{tab:energy-rms}.
Our calculations reproduce well the experimental data of the ground-state
energy~\cite{Purcell10}.
On the other hand, some deviation of the calculated root-mean-square radii
from the experimental data is found.
However, the difference does not affect the reaction analysis as shown below.
\begin{table}[htbp]
 \centering
 \caption{Parameters of Gaussian basis}
 \begin{tabular}{ccccccc} \hline
  c & $i_{\rm max}$ & $x_0$[fm] & $x_{\rm max}$[fm] &
  $j_{\rm max}$ & $y_0$[fm] & $y_{\rm max}$[fm] \\ 
  \hline \hline
  1, 2 & 12 & 0.1 & 20.0 & 12 & 0.1 & 20.0 \\
  3 & 12 & 0.1 & 20.0 & 12 & 0.1 & 20.0 \\
  \hline    
 \end{tabular}
 \label{tab:parameters-GEM}
\end{table}
\begin{table}[htbp]
 \centering
 \caption{
 Ground-state energies and root-mean-square radii of $t$ and $^3$He.
 The experimental data are taken from Ref.~\cite{Purcell10}.
 }
 \begin{tabular}{ccccc} 
  \hline
  & \hspace{5em} Cal. & & \hspace{5em} Exp. & \\
  \hline
  & $\varepsilon_{0}$ [MeV] & $r_{\rm rms}$ [fm] & $\varepsilon_{0}$ [MeV]
  & $r_{\rm rms}$ [fm] \\
  \hline \hline
  $t$ & $-8.45$ & 1.68 & $-8.48$ & 1.84 \\
  $^3$He & $-7.77$ & 1.70 & $-7.71$ & 1.99 \\
  \hline    
 \end{tabular}
 \label{tab:energy-rms}
\end{table}

\begin{figure}[tpb]
 \centering
 \includegraphics[scale=0.8]{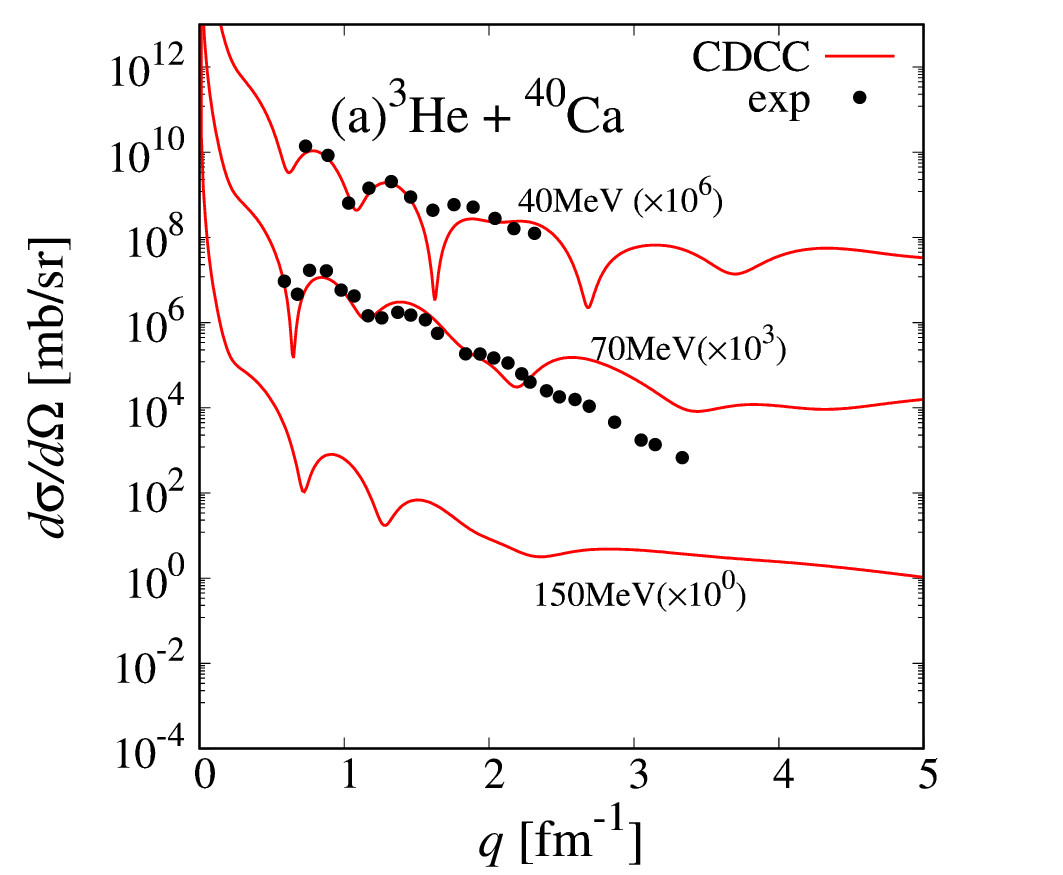}
 \includegraphics[scale=0.8]{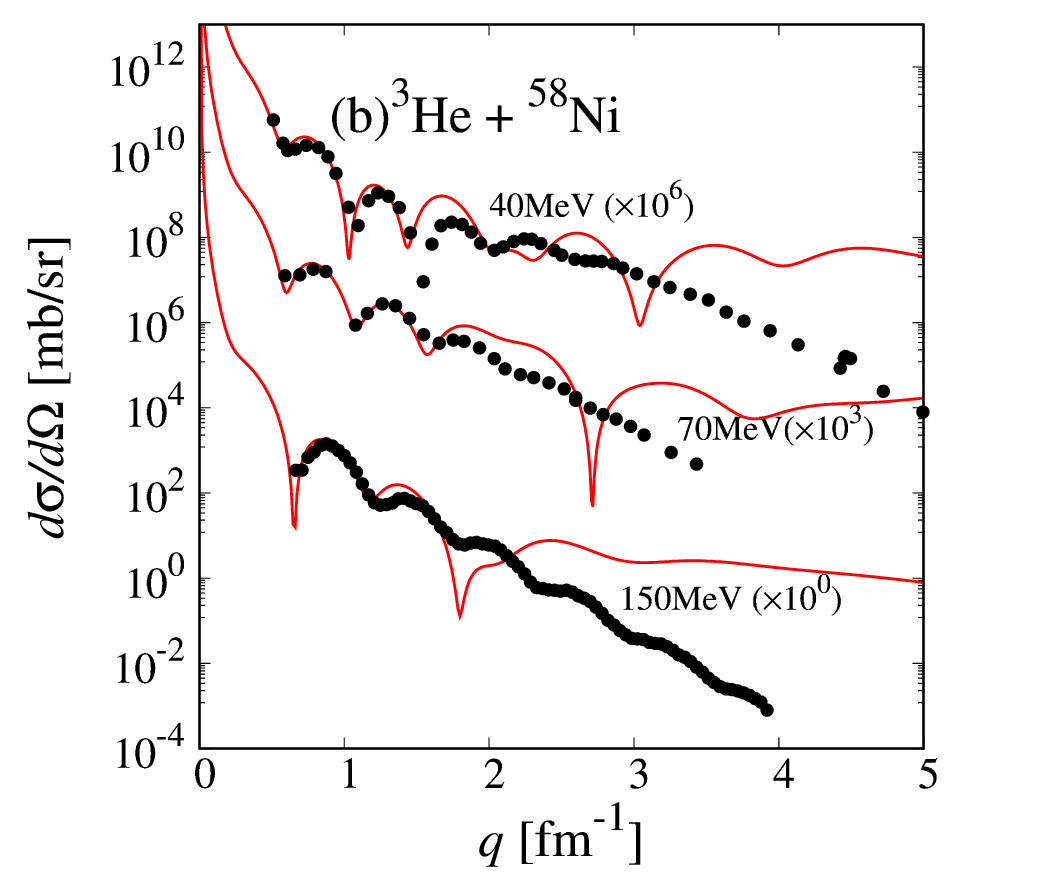}
 \includegraphics[scale=0.8]{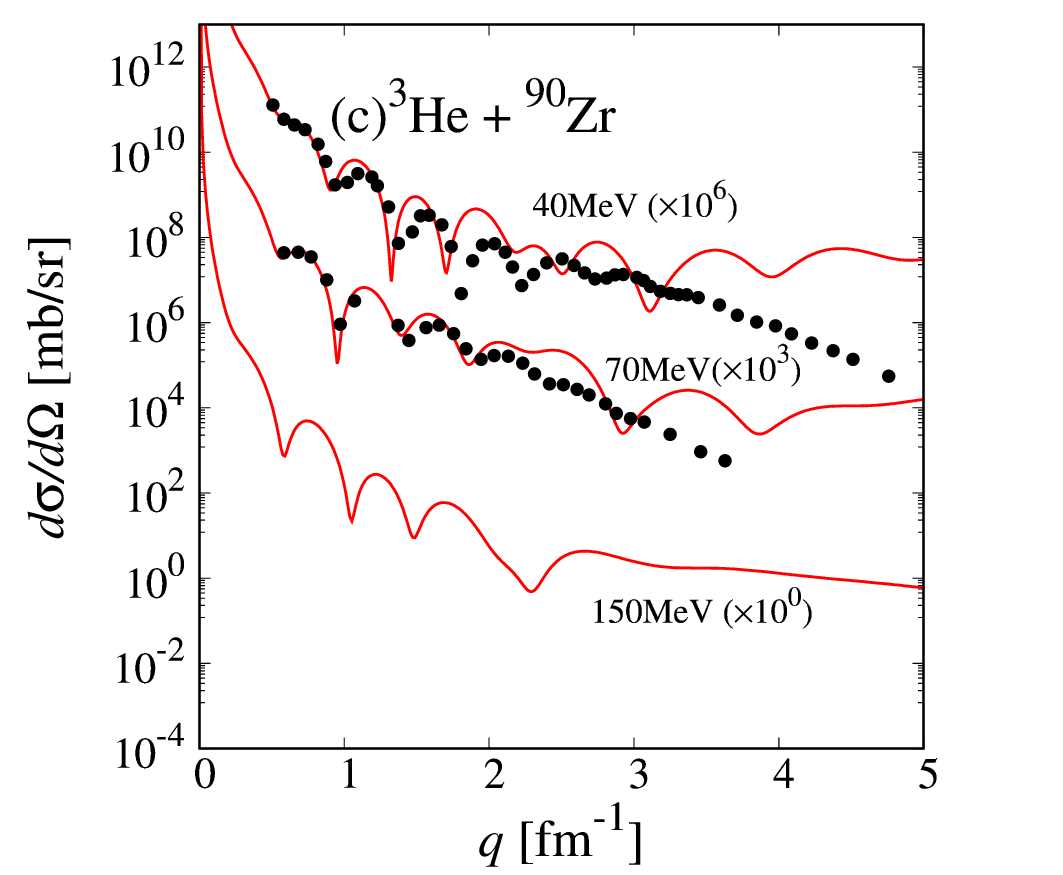}
 \caption{
 Transferred momentum distributions of elastic cross sections
 of $^3$He with (a) $^{40}$Ca, (b) $^{58}$Ni, and (c) $^{90}$Zr targets
 at 40, 70, and 150 MeV/nucleon.
 The experimental data are taken from 
 Refs.~\cite{Tabor82,Willis73,Hyakutake80,Kamiya03}.}
 \label{fig:elastic}
\end{figure}

In order to confirm the validity of our three-body model,
we analyze elastic scattering of $^3$He off $^{40}$Ca, $^{58}$Ni, and $^{90}$Zr.
In the model space, we include continuum states up to the internal energy 
$\varepsilon$ of 30 MeV for $I^{\pi}=$ $0^+$, $1^-$, and $2^+$ states of the projectile.
$U_i$ in Eq.~\eqref{eq:4body-eq} is constructed by folding the Melbourne $g$ 
matrix~\cite{Amos00} with the target density~\cite{Minomo10}.
Figure~\ref{fig:elastic} shows the elastic cross sections of $^3$He at 40, 70, and 150 MeV/nucleon, 
as a function of the transferred momentum $q$.
The experimental data of the cross sections denoted by the dots are taken from 
Ref.~\cite{Tabor82,Willis73,Hyakutake80,Kamiya03}.
The solid lines represent the results of the E-CDCC calculation.
It is found that the E-CDCC results reproduce the experimental data
in the small $q$ region, in which the cross section is large.
Therefore, the three-body model for $^3$He adopted in this study is expected to work well.
\subsection{Breakup properties of $t$ and $^3$He}
\begin{figure*}[htbp]
 \begin{center}
  \includegraphics[scale=0.65]{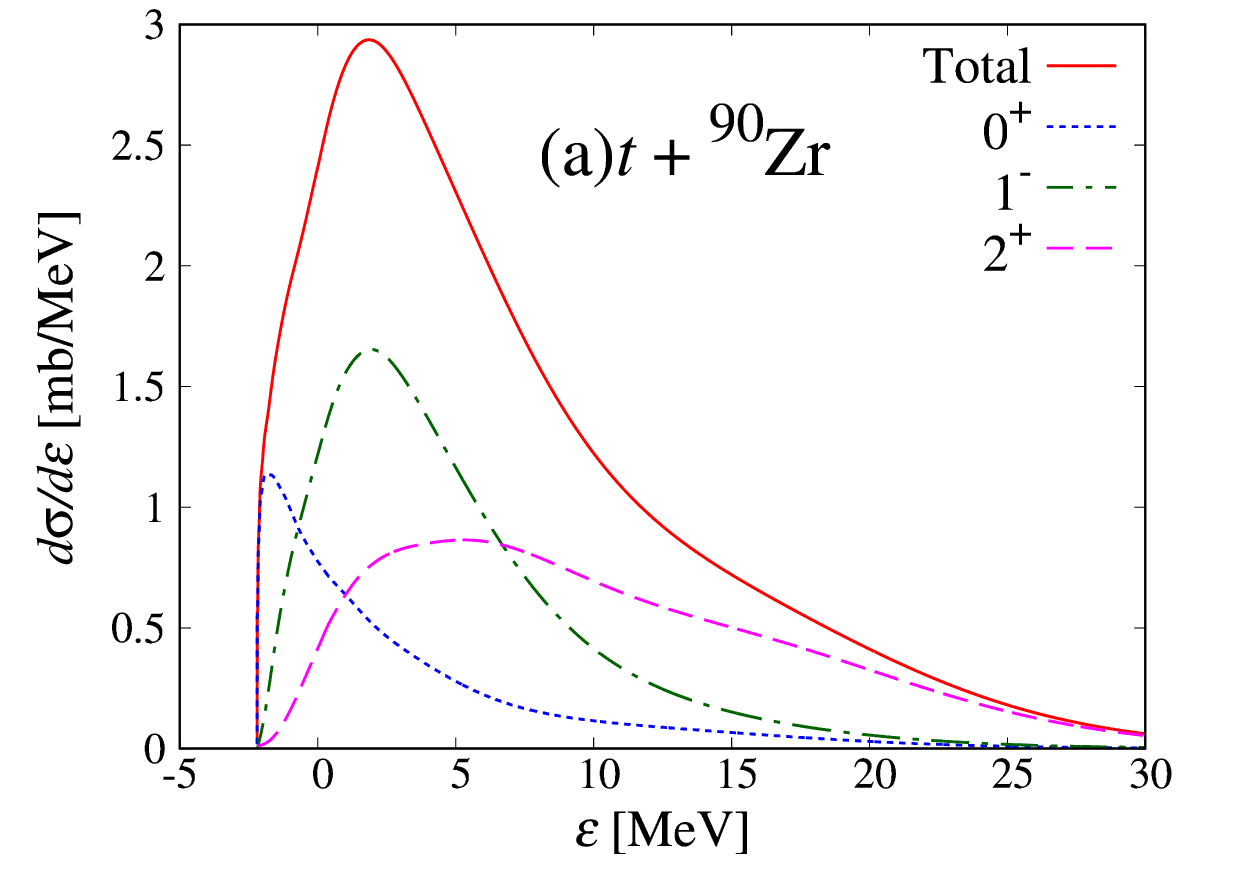}
  \includegraphics[scale=0.65]{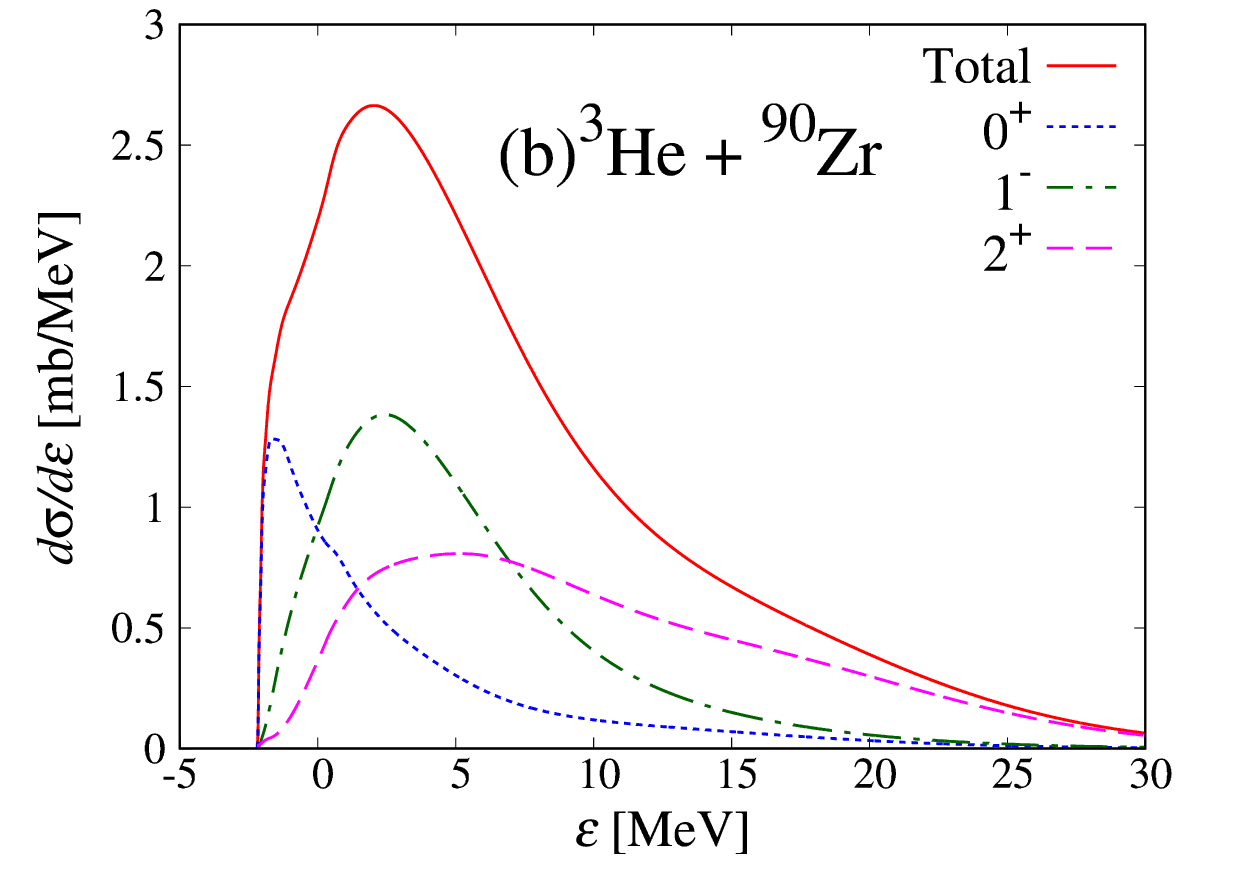}
  \caption{(a) Breakup cross section of $t$ with a $^{90}$Zr target 
  at $E$ = 150 MeV/nucleon. (b) Same as (a) but of $^3$He.}
  \label{fig:dbu-3He90Zr}
 \end{center}
\end{figure*}
We investigate the breakup effects of $t$ and $^3$He on the breakup energy spectra 
and compare them with those of $d$.
We adopt the method proposed in Ref.~\cite{Matsumoto10} to obtain a smooth breakup spectrum.
The model space is the same as in the calculation of the elastic scattering.

First, we show the breakup cross sections of the $t$ ($^3$He) + $^{90}$Zr reaction 
at 150 MeV/nucleon in Fig.~\ref{fig:dbu-3He90Zr}(a) (Fig.~\ref{fig:dbu-3He90Zr}(b)).
$\varepsilon$ = $-2.2$ and $0$ MeV correspond to the thresholds of the $d+n$ ($d+p$)
and $n+n+p$ ($p+p+n$) channels for $t$ ($^3$He), respectively.
One can see that the behaviors of the breakup cross section of $t$ and $^3$He are almost the same.
This can be understood from the fact that the strength of the electric dipole transition,
which mainly contributes to Coulomb breakup reactions, for $t$ is the same as for $^3$He;
details are found in Appendix.
The similar behavior of the cross section between $t$ and $^3$He is also confirmed 
in other reaction systems. 
Thus, in what follows, we will concentrate on the results of $^3$He.

\begin{figure*}[htbp]
 \begin{center}
  \includegraphics[scale=0.9]{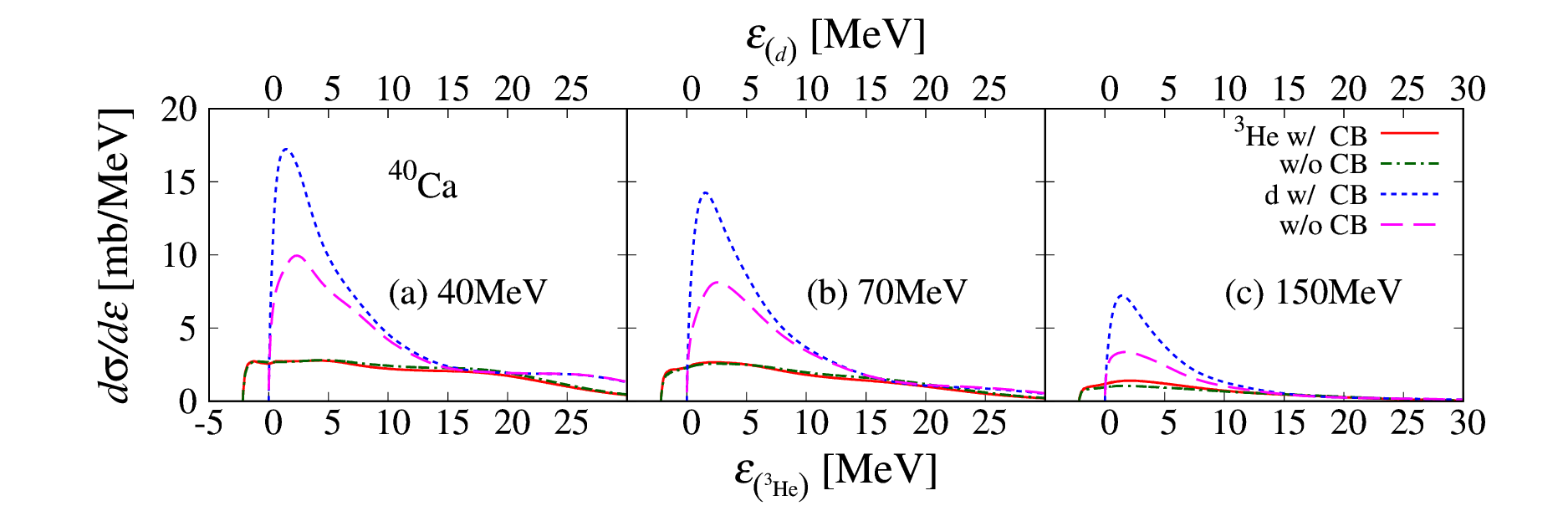}\\
  \caption{The energy spectra of breakup cross sections of $^3$He and $d$
  with a $^{40}$Ca target at (a) 40 MeV/nucleon, (b) 70 MeV/nucleon, and (c) 150 MeV/nucleon.
  The upper (lower) horizontal axis shows the breakup energy of $d$ ($^3$He) 
  regarding the $p+n$ ($p+p+n$) threshold.}
  \label{fig:dbu-3He40Ca-d-CB}
 \end{center}
\end{figure*}
\begin{figure*}[htbp]
 \begin{center}
  \includegraphics[scale=0.9]{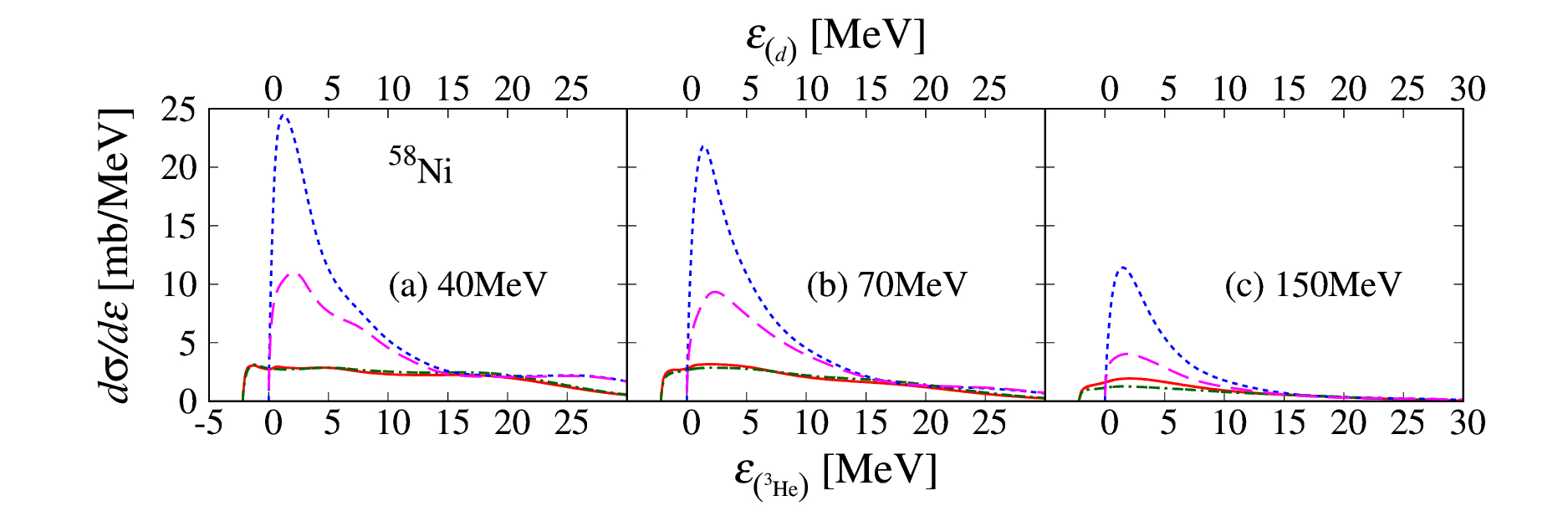}\\
  \caption{Same as in Fig.~\ref{fig:dbu-3He40Ca-d-CB} but with a $^{58}$Ni target.}
  \label{fig:dbu-3He58Ni-d-CB}
 \end{center}
\end{figure*}
\begin{figure*}[htbp]
 \begin{center}
  \includegraphics[scale=0.9]{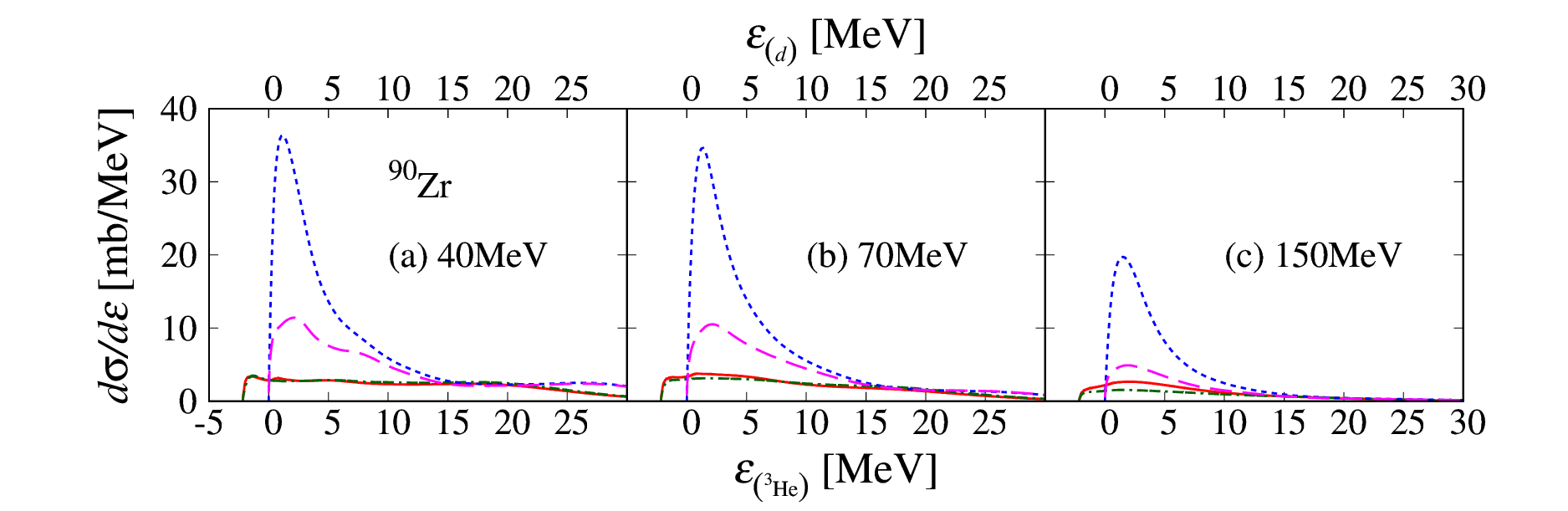}
  \caption{Same as in Fig.~\ref{fig:dbu-3He40Ca-d-CB} but with a $^{90}$Zr target.}
  \label{fig:dbu-3He90Zr-d-CB}
 \end{center}
\end{figure*}
Next, we compare systematically the breakup cross sections of $^3$He and $d$.
For the E-CDCC calculation of the $d$ breakup reaction,
the optical potentials are constructed in the same manner as of $t$ and $^3$He.
We include continuum states of $d$ up to $\varepsilon=$ 30 MeV 
for $I^{\pi}$ = $0^+$, $1^-$, and $2^+$ states; 
the spin of each nucleon is neglected, as in the description of $t$ and $^3$He.
The solid and dotted lines in Figs.~\ref{fig:dbu-3He40Ca-d-CB}, 
\ref{fig:dbu-3He58Ni-d-CB}, and \ref{fig:dbu-3He90Zr-d-CB} represent the results 
for $^3$He and $d$, respectively.
The dot-dashed (dashed) line corresponds to the result with only 
the nuclear breakup of $^3$He ($d$).
Although the effective charge of $^3$He is 2/3$e$, which is larger than 1/2$e$ of $d$,
the Coulomb breakup of $^3$He is negligible compared to that of $d$
because of the large binding energy of $^3$He.
We show the total breakup cross section of $^3$He and $d$ in TABLE~\ref{tab:total-bux-CB}.
The results for $^3$He are found to be about one-third of those for $d$ in all the cases.
\begin{table}[tbp]
 \tblcaption{
 Total breakup cross section (in mb) of $^3$He and $d$ for the present reaction systems.
 $E$ represents the incident energy per nucleon.
 }
 \label{tab:total-bux-CB}
 \begin{tabular}{cp{3em} cp{6em} cp{6em} cp{6em} cp{6em} cp{6em} cp{6em}}
  \hline
  &\hspace{0.5em} $E$ &\multicolumn{2}{c}{40 MeV}
  &\multicolumn{2}{c}{70 MeV}
  &\multicolumn{2}{c}{150 MeV}
  \\
  \hline
  &\multicolumn{1}{c}{}
  &\multicolumn{1}{c}{~$\sigma^{(d)}_{\rm BU}$}
  &\multicolumn{1}{c}{~$\sigma^{(^3{\rm He})}_{\rm BU}$}          
  &\multicolumn{1}{c}{~$\sigma^{(d)}_{\rm BU}$}
  &\multicolumn{1}{c}{~$\sigma^{(^3{\rm He})}_{\rm BU}$}
  &\multicolumn{1}{c}{~$\sigma^{(d)}_{\rm BU}$} 
  &\multicolumn{1}{c}{~$\sigma^{(^3{\rm He})}_{\rm BU}$}\\
  \hline
  &\multicolumn{1}{c}{$^{40}$Ca}
  &\multicolumn{1}{c}{~148}
  &\multicolumn{1}{c}{~62}
  &\multicolumn{1}{c}{~114}
  &\multicolumn{1}{c}{~48}
  &\multicolumn{1}{c}{~47}
  &\multicolumn{1}{c}{~20}
  \\
  &\multicolumn{1}{c}{$^{58}$Ni}
  &\multicolumn{1}{c}{~181} 
  &\multicolumn{1}{c}{~68}
  &\multicolumn{1}{c}{~150}
  &\multicolumn{1}{c}{~56}
  &\multicolumn{1}{c}{~69}
  &\multicolumn{1}{c}{~26}
  \\
  &\multicolumn{1}{c}{$^{90}$Zr}
  &\multicolumn{1}{c}{~228} 
  &\multicolumn{1}{c}{~71}
  &\multicolumn{1}{c}{~205}
  &\multicolumn{1}{c}{~64}
  &\multicolumn{1}{c}{~108}
  &\multicolumn{1}{c}{~33}
  \\
  \hline
 \end{tabular}       
\end{table}
\begin{figure}[tbp]
 \begin{center}
  \includegraphics[scale=0.65]{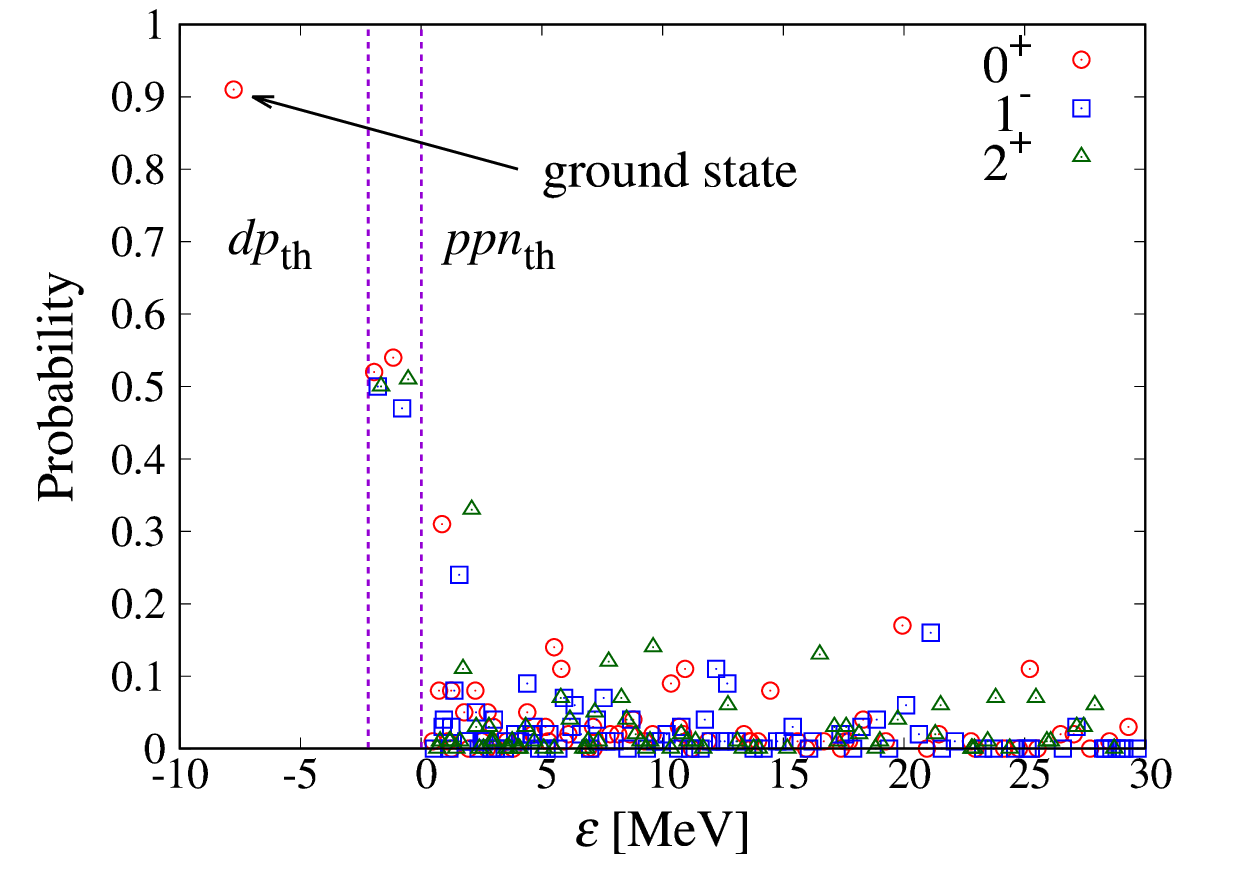}
  \caption{Probabilities of the existence of $d$ 
  in the ground and discretized states of $^3$He.
  The left and right vertical dotted lines represent
  the thresholds of the $d+p$ and $p+p+n$ channels, respectively.
  }
  \label{fig:overlap_3He}
 \end{center}
\end{figure}
Next, we investigate the mechanism of the $^3$He breakup reaction, i.e., 
the decomposition of the breakup channels into the following two:
\begin{eqnarray*}
 ^3{\rm He} &\rightarrow& d + p , \\
 ^3{\rm He} &\rightarrow& p + p + n .
\end{eqnarray*}
For this purpose, we use the P-separation method proposed in Ref.~\cite{Watanabe21}.
In this method, 
the probability $P_\gamma$ of the existence of $d$ in $\Phi_\gamma$ is defined by
\begin{eqnarray}
 \label{eq:probability}
 P_\gamma = 
  \int 
  \braket{\Phi_{\gamma}(\bm{x},\bm{y})|\chi_d(\bm{x}_1)}_{\bm{x}_1}
  \braket{\chi_d(\bm{x}_1)|\Phi_{\gamma}(\bm{x},\bm{y})}_{\bm{x}_1} 
  d\bm{y}_{1} ,
  \nonumber \\
\end{eqnarray}
where $\chi_d$ is the wave function of $d$.
Then, by using $P_\gamma$, the $d+p$ and $p+p+n$ channel contributions to
the total breakup cross section can be obtained as follows:
\begin{eqnarray}
 \sigma_{d+p} &\equiv& \sum_{\gamma} P_{\gamma} \sigma_{\gamma} , \\
 \sigma_{p+p+n} &\equiv& \sum_{\gamma} (1-P_{\gamma}) \sigma_{\gamma}.
\end{eqnarray}
Here, $\sigma_{\gamma}$ is the breakup cross section
to the discretized state $\Phi_\gamma$.

\begin{figure}[tbp]
 \begin{center}
  \includegraphics[scale=1.2]{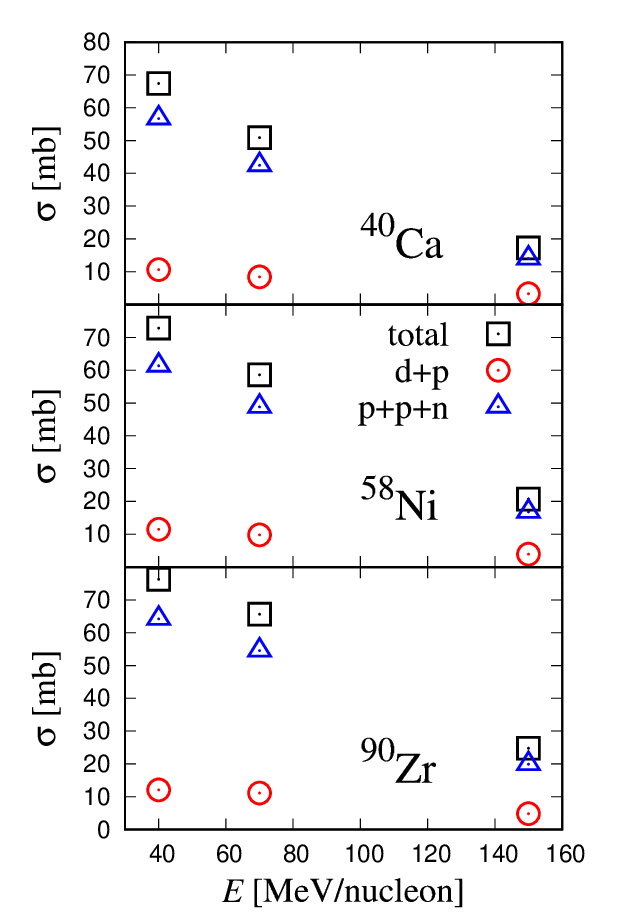}
  \caption{Total breakup cross sections for $^3$He breakup reactions 
  separated into the $d+p$ and $p+p+n$ channels by using the P-separation method.}
  \label{fig:p-separation}
 \end{center}
\end{figure}
Figure~\ref{fig:overlap_3He} shows the results of $P_\gamma$ for each state of $^3$He.
For the ground state, $P_{0}$ $\approx$ 0.9 is obtained, 
which is consistent with the value 90\%~\cite{Brida11} obtained with
the {\it ab initio} quantum Monte Carlo calculation.
The left and right vertical dotted lines in Fig.~\ref{fig:overlap_3He} represent
the thresholds of the $d+p$ and $p+p+n$ channels, respectively.
$P_\gamma$ for the continuum states between the two thresholds are found to be about 0.5.
For other states, $P_\gamma$ are mostly smaller than 0.3.
This result indicates that the $p+p+n$ channel contribution is dominant 
in the total breakup cross section of $^3$He.

Figure~\ref{fig:p-separation} shows the results of $\sigma_{d+p}$ and $\sigma_{p+p+n}$.
In this calculation, we do not include the Coulomb breakup because its contribution 
is negligible as mentioned above.
One sees that the contribution of the $p+p+n$ three-body breakup is about five times
as large as that of the $d+p$ two-body breakup in all of the three reaction systems.
This behavior can be understood from the larger three-body phase volume.
It should be noted, however, that this is not always the case. 
In Ref.~\cite{Watanabe21}, the authors found with a similar approach that, 
for the breakup of $^6$Li, 
the $d+{}^{4}$He two-body channel is more important than the $p+n+{}^{4}$He one, 
despite that the latter has a larger phase volume. 
Further investigations are needed to clarify the relation between the cross sections 
and the sizes of the phase volume for two-body and three-body breakup processes. 
In any case, the results in Fig.~\ref{fig:p-separation}
suggest that the $^3$He breakup reaction should be described as a four-body breakup reaction.

\subsection{Four-body and three-body reactions}

\begin{figure}[tbp]
 \begin{center}
  \includegraphics[scale=0.7]{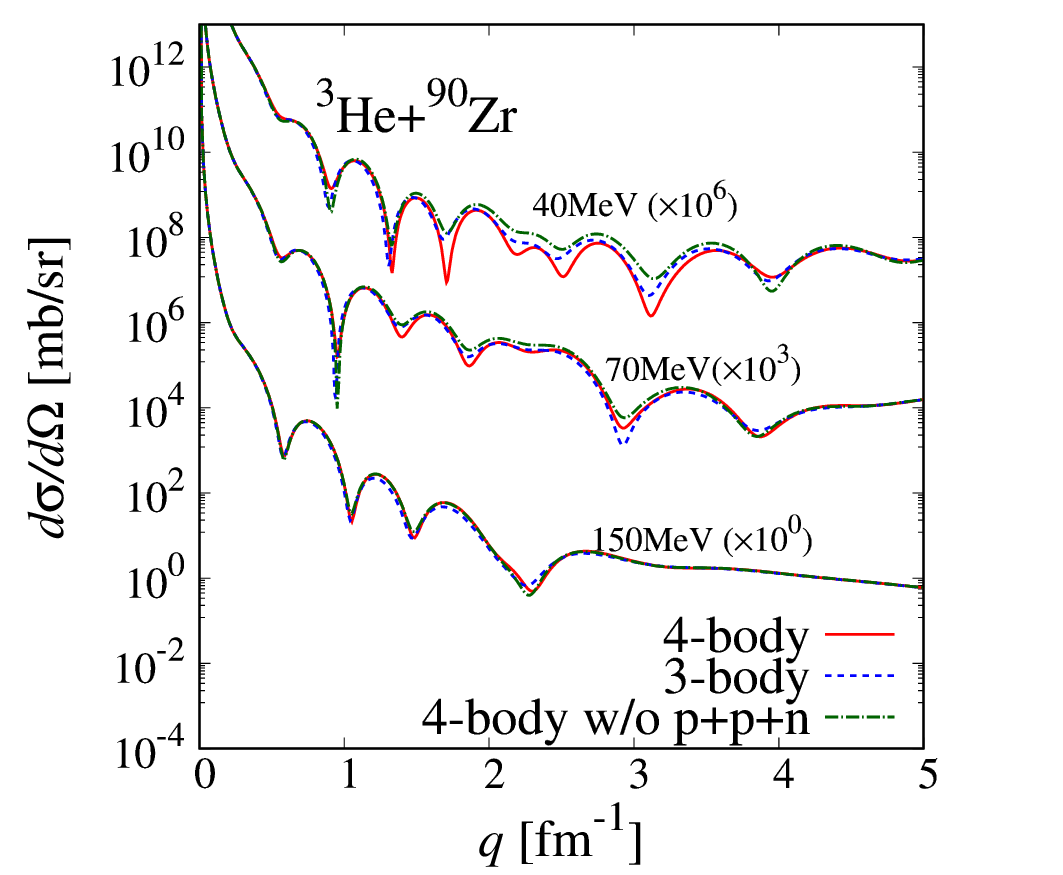}
  \caption{Transferred momentum distributions of elastic cross section off $^{90}$Zr.
  The energies in the panels represent the incident energies per nucleon.
  }
  \label{fig:ela-4body-3body}
 \end{center}
\end{figure}

In the present study, we have analyzed the $^3$He reaction with a $p+p+n+{\rm T}$
four-body model, whereas in Ref.~\cite{Iseri86}, 
it was investigated with a $d+p+{\rm T}$ three-body model.
We investigate the difference between the two reaction models.
To describe $^3$He as a two-body model, 
we use the same potential between $d$ and $p$ as in Ref.~\cite{Iseri86}.
While the previous study adopted a phenomenological potential for the optical potential 
between $d$ and T, 
we use the following folding-model potential:
\begin{eqnarray}
 U_d = \braket{\chi_d | U_p + U_n |\chi_d} .
\end{eqnarray}
It should be noted that the three-body calculation using $U_d$ does not include breakup of $d$.
The optical potential between $p$ and T is the same as used in the four-body calculation.
The solid and dotted lines in Fig.~\ref{fig:ela-4body-3body} represent the results of the E-CDCC
calculation with the four-body and three-body reaction models with a $^{90}$Zr target,
respectively, as a function of $q$. 
We have included only the nuclear breakup in this calculation.
Although some differences are found around the dips at low incident energy,
the shapes of the oscillations are almost the same.
The difference of the depth around the dips is considered to come from the effects of 
the $p+p+n$ channel.
To discuss this in detail,
we perform the four-body E-CDCC calculation including only the ground state and the $d+p$
continuum states located between the two vertical dotted lines in Fig.~\ref{fig:overlap_3He}.
The dot-dashed lines thus obtained are close to the results of the three-body calculation. 
This confirms the slight effect of the $p+p+n$ channel on the elastic scattering.

\begin{figure}[tbp]
 \begin{center}
  \includegraphics[scale=0.6]{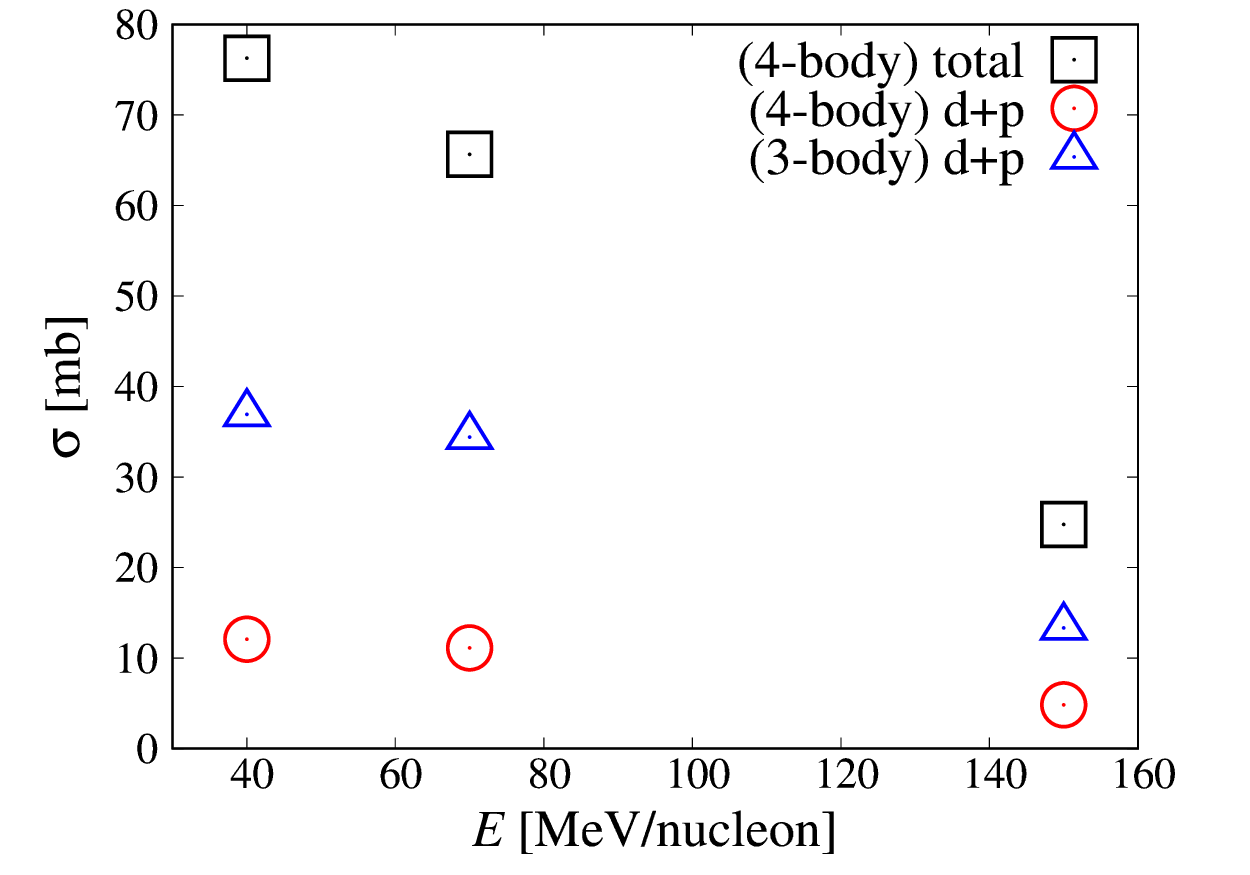}
  \caption{Comparison between the breakup cross sections obtained by four- and three-body 
  E-CDCC with a $^{90}$Zr target.}
  \label{fig:4body-3body}
 \end{center}
\end{figure}
Figure~\ref{fig:4body-3body} shows the comparison of the total breakup cross section 
with a $^{90}$Zr target calculated with four- and three-body E-CDCC.
The squares and circles are the same as in Fig.~\ref{fig:p-separation},
whereas the triangles represent the cross sections calculated with three-body E-CDCC.
The total breakup cross sections obtained with the four-body calculation 
are two times as large as those with the three-body calculation. 
This difference can be basically understood from the significant 
contribution of the $p+p+n$ channel, which is missing in three-body E-CDCC,  
in the $^3$He breakup reaction.
In addition, 
it is suggested by the difference between the triangles and circles that
the $d+p$ two-body breakup process is suppressed in the four-body calculation, 
probably because of the coupling between the $d+p$ and $p+p+n$ channels.

\section{Summary}

We have investigated the $t$ and $^3$He breakup reactions with four-body E-CDCC.
We treated $t$ and $^3$He as three-nucleon systems.
The elastic scattering cross section data of $t$ and $^3$He are reproduced
well by the present framework.
For the analysis of breakup reactions,
we take into account the nuclear and Coulomb breakup in the E-CDCC calculations.
The breakup cross sections of $t$ and $^3$He are almost the same for the reaction systems considered.
The Coulomb breakup of $^3$He is found to be negligibly small, 
and the total breakup cross section of $^3$He is about one-third of that of $d$.

In addition,
we applied the P-separation method to the investigation of the final channels of the $^3$He 
breakup reaction and showed that the contribution of the $p+p+n$ channel is dominant.
We have further investigated the difference between the four-body E-CDCC calculation 
and three-body one; in the latter, $^3$He is described as a $d+p$ system.
These two models are found to give almost the same result for the elastic scattering.
For the breakup reaction,
the total breakup cross section calculated with four-body E-CDCC is as twice as that with three-body E-CDCC.
Thus, we conclude that the $t$ and $^3$He breakup reactions should be treated as 
the four-body reaction.

\section*{Acknowledgements} 
This work is supported in part by Grant-in-Aid for Scientific Research
(No.\ JP22K14043, No.\ JP21H00125, and No.\ JP21H04975)
from Japan Society for the Promotion of Science (JSPS).

\section*{Appendix}
\begin{figure}[bp]
 \centering
 \includegraphics[scale=0.3]{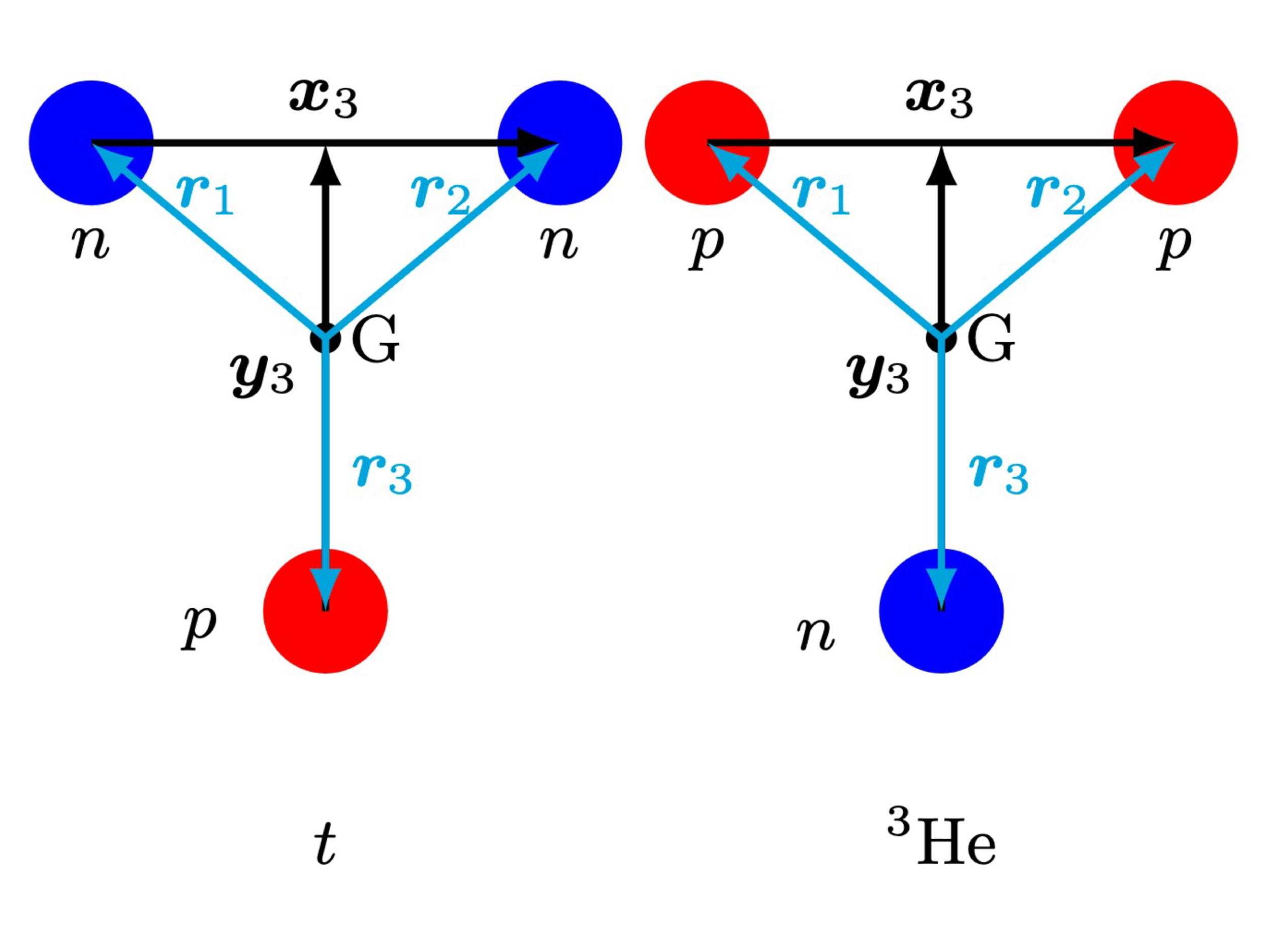}
 \caption{
 The coordinate from c.m. to each particle represented with the Jacobi coordinate.
 G means the c.m. of the $t$ and $^3$He.
 }
 \label{fig:E1strength}
\end{figure}
The electric dipole (E1) transition operator is defined as 
\begin{eqnarray}
 \label{eq:E1strength}
 D_{\mu} = \sum_{i=1}^{3} 
  \left(\frac{1}{2}-\tau_{iz}\right)e r_{i} Y_{1\mu}(\Omega_{r_{i}}) ,
\end{eqnarray}
where $\tau_{iz}$ is the $z$ component of the isospin operator.
$r_i$ means the coordinate from c.m. to each particle as shown in Fig.~\ref{fig:E1strength},
and can be represented as follows by using the 
Jacobi coordinate $\{\bm{x}_{3},\bm{y}_{3}\}$:
\begin{eqnarray}
 \bm{r}_{1} &=& -\frac{1}{2}\bm{x}_{3} + \frac{1}{3} \bm{y}_{3} ,
  \nonumber \\
 \bm{r}_{2} &=& \frac{1}{2}\bm{x}_{3} + \frac{1}{3} \bm{y}_{3} ,
  \\
 \bm{r}_{3} &=& - \frac{2}{3} \bm{y}_{3} .
  \nonumber
\end{eqnarray}
Using this relation, the spherical harmonics is written as
\begin{eqnarray}
 \label{eq:spherical}
 r_{1} Y_{1\mu}(\Omega_{r_1})
  &=&
  -\frac{1}{2} x_{3} Y_{1\mu}(\Omega_{x_3})
  +\frac{1}{3} y_{3} Y_{1\mu}(\Omega_{y_3}) ,
  \nonumber \\
 r_{2} Y_{1\mu}(\Omega_{r_2})
  &=&
  \frac{1}{2} x_{3} Y_{1\mu}(\Omega_{x_3})
  +\frac{1}{3} y_{3} Y_{1\mu}(\Omega_{y_3}) ,
  \\
 r_{3} Y_{1\mu}(\Omega_{r_3})
  &=&
  -\frac{2}{3} y_{3} Y_{1\mu}(\Omega_{y_3}) .
  \nonumber
\end{eqnarray}
Inserting Eq.~\eqref{eq:spherical} to Eq.~\eqref{eq:E1strength},
we can obtain
\begin{eqnarray}
 \label{eq:E1-t}
 D_{\mu} = -\frac{2}{3} e y_{3} Y_{1\mu}(\Omega_{y_{3}})
\end{eqnarray}
for $t$ and 
\begin{eqnarray}
 \label{eq:E1-3He}
 D_{\mu} = \frac{2}{3} e y_{3} Y_{1\mu}(\Omega_{y_{3}})
\end{eqnarray}
for $^3$He.
Thus, $t$ and $^3$He have the same E1 effective charge.

\bibliography{./ref}

\end{document}